\documentclass[useAMS,usenatbib]{mn2e}
\usepackage{amssymb}
\usepackage{amsmath}
\usepackage{graphicx}
\usepackage{rotating}
\usepackage{lscape}
\usepackage{array}
\def\hi{H\,{\sc i}}

\def\deg{$^{\circ}$}
\def\kms{km~s$^{-1}$}
\def\msun{M$_{\odot}$}
\def\Ha{H$\sf \alpha$}
\def\micron{$\mu$m}

\def\msunppc{M$_{\odot}$~pc$^{-3}$}

\long\def\symbolfootnote[#1]#2{\begingroup%
\def\thefootnote{\fnsymbol{footnote}}\footnote[#1]{#2}\endgroup} 

\def\aj{AJ}%
\def\apj{ApJ}%
\def\apjl{ApJ}%
\def\apjs{ApJS}%
\def\aap{A\&A}%
\def\aaps{A\&AS}%
\def\mnras{MNRAS}%
\def\pasa{PASA}%
\def\nat{Nature}%
\title[\hi\ synthesis observations of NGC~1705]{\hi\ synthesis observations of the blue compact dwarf NGC~1705}
\author[Elson~et~al.]{E.~C.~Elson$^{1,2}$\thanks{E-mail:
elson.e.c@gmail.com (ECE), blok@astron.nl (WJGdeB), kraan@ast.uct.ac.za (RCK-K)}, W. J. G. de Blok$^{1,3}$ and R. C. Kraan-Korteweg$^{1}$\\
$^{1}$Astrophysics, Cosmology and Gravity Centre (ACGC), Department of Astronomy, University of Cape Town,\\ Private Bag X3,
Rondebosch 7701, South Africa\\
$^{2}$International Centre for Radio Astronomy Research, The University of Western Australia, M468,\\ 35 Stirling Highway, Crawley, Western Australia, 6009, Australia\\
$^{3}$Netherlands Institute for Radio Astronomy (ASTRON), Postbus 2, 7990 AA Dwingeloo, the Netherlands}

\begin{document}


\pagerange{\pageref{firstpage}--\pageref{lastpage}} \pubyear{2002}

\maketitle

\label{firstpage}
\begin{abstract}
Australia Telescope Compact Array \hi-line observations of the nearby dwarf galaxy NGC~1705 are presented.  The data are used to trace the gravitational potential of the galaxy out to several stellar disc scale lengths.  A rotation curve is derived for the system and used to generate mass models.  Dark matter dominates the gravitational potential at nearly all galactocentric radii.  NFW and pseudo-isothermal sphere halo parameterisations allowing for good reproductions of the observations.  The models suggest NGC~1705 to have a dark matter halo that it much denser and more compact than previously thought.   
\end{abstract}


\section{Introduction}
Studies carried out over the past two decades have shown dwarf galaxies in the local Universe to be dark-matter-dominated systems \citep[e.g.][and references therein]{carignan_beaulieu_DDO154_1989,cote_carignan_sancisi_1991,broeils_phd,meurer2,deblok_mcgaugh_1996, swaters_thesis, THINGS_deblok}.  \citet{swaters_thesis} studied the dark matter properties of a large representative sample of nearby dwarf galaxies as part of the Westerbork \hi\ Survey of Spiral and Irregular Galaxies \citep[WHISP, ][]{WHISP_swaters}.  Confirming the results of previous studies \citep[e.g.][]{broeils_phd}, he showed the rotation curves in his sample not to decline at large radii, thereby providing evidence for large unseen mass components.  



A nearby dwarf galaxy with an unusually large \hi\ disc extending out to several optical scale lengths is NGC~1705.  The position of the optical centre of the system is $\alpha_{2000}$~=~04$^{\mathrm{h}}$~54$^{\mathrm{m}}$~14.$^{\mathrm{s}}$4, $\delta_{2000}$~=~$-53^{\circ}$~21$'$~38$''$ \citep{meurer_SINGG_96} and its distance $D=5.1\pm0.6$~Mpc \citep[][]{tosi_et_al_2001}.  Given the proximity of NGC~1705, the \hi\ disc serves as a useful tracer of the system's gravitational potential.  \citet{meurer_1705_2} were the first group of investigators to carry out \hi-line observations of NGC~1705 and to produce a mass model of its rotation curve.  They found the galaxy to be dark-matter-dominated at nearly all radii with a central dark matter density $\rho_0\sim 0.1$~\msun~pc$^{-3}$ -- very high compared to other dwarfs and late-type systems.  In this work we present new \hi-line observations of NGC~1705 obtained with the Australia Telescope Compact Array.  We combine these new data with \emph{Spitzer} and GALEX imaging to derive mass models for the galaxy.
 
Besides being famous for its dark matter content, NGC~1705 is also well-known for hosting one of the most powerful starbursts (relative to its mass) in the local Universe.  This super star cluster, first described by \citet{sandage_1978} and later reported on in detail by \citet{melnick_1985}, contributes almost half of the total ultra-violet luminosity of the galaxy \citep{meurer_1705_2}.  The system has a $B$-band absolute magnitude of $M_B = -15.6 \pm 0.2$ \citep{marlowe_1999}.  A detailed study of the star formation activity of NGC~1705 was carried out by \citet{elson_SF_models_2012}.  The authors find star formation to be regulated by the \hi\ kinematics as well as the central concentration of dark matter.   

NGC~1705 is located in a low-density environment. No nearby companion galaxies are seen in our \hi\ data cubes which span a volume $\sim 20$~kpc~$\times~20$~kpc~$\times~4$~Mpc.  The NASA/IPAC Extragalactic data base was used to search for objects within 10\deg\ ($\sim 0.9$~Mpc) of NGC~1705 spanning the velocity range 550~--~700~\kms\ ($\sim 2$~Mpc).  The only catalogued objects in this volume besides NGC~1705 are two ultra compact dwarf galaxies \citep{hilker,drinkwater, phillipps} separated more than 0.5~Mpc from NGC 1705.  Given that these systems are less luminous ($-13< M_B < -11$) than any of the known compact dwarf galaxies, any interaction effects they might have on NGC~1705 will be negligible.  Finally, we also searched a HIPASS cube of 2\deg~$\times~2$\deg~$\times~300$~\kms\ centred on the galaxy, finding no evidence for nearby companions.

The structure of this paper is as follows.  The new \hi\ observations are presented in Section~\ref{HI_data} and the data products in Section~\ref{HI_properties}.  In Section~\ref{blow_out} we discuss the possibility of a galactic wind blow out for the galaxy.  A rotation curve is shown in Section~\ref{rotation_curve_sec} along with the corresponding mass models in Section~\ref{mm}.  Our summary and conclusions are presented in Section~\ref{1705_HI_summary}. 

\section{\hi\ observations and data reduction}\label{HI_data}
\subsection{Data acquisition}
In this work we utilise the \hi\ data set presented by \citet{elson_SF_models_2012}.  NGC~1705 was observed between 22~November~2006 and 17~March~2007 with six different Australia Telescope Compact Array (ATCA) configurations using all six antennas  (project number C1629).  Table \ref{1705_ATCA_table} lists the details of each observing run.  A single run consisted of a primary calibrator observation, regular secondary calibrator observations and source observations.  PKS~B1934-630 and PKS~B0407-658 were used as primary and secondary calibrators, respectively, for all runs.   Most runs were approximately 11 hours long.  A single run was carried out in each of the EW352, 750D, and 1.5B antenna configurations whereas four runs were carried out in the 6A configuration in order to boost high-resolution sensitivity to low-surface-brightness emission.  No mosaicking was required and the telescope pointing centres were set to the optical centre position of NGC~1705.  The correlator was set to use 512 channels with a bandwidth of 8 MHz centred at 1418 MHz.  The resulting velocity range is $-126$~\kms\ to 1561~\kms.
  
\begin{table*}
\begin{center}
\caption{Summary of NGC~1705 observing set-ups.}
\label{1705_ATCA_table}
\begin{tabular}{ccccc}
\hline
\hline
\\
Configuration & Date  &  Start & End & Duration\\ 
         & (yy-mm-dd) & (hh-mm-ss) & (hh-mm-ss) & (h)\\ 
         1	& 2 	& 3 	  & 4 	  & 5\\ 
\\
\hline
\\
 EW352 & 2006-10-22 & 11:18:27 & 22:59:16 & 11.68\\
\\
  750D  & 2007-03-17 & 05:24:35 & 12:51:46 & ~~7.59\\ 
\\
 1.5B  & 2006-11-24 & 07:42:00 & 18:57:17 & 11.25\\
\\
 1.5B   & 2006-11-25 & 07:27:17 & 18:57:15 & 11.50\\
\\
6A   & 2007-02-13 & 02:11:55 & 14:11:42 & 12.00\\
\\
 6A    & 2007-02-14 & 02:08:38 &  12:34:42 & 10.43\\
\\
 6A    & 2007-02-17 & 01:29:35 &  11:05:57 & ~~9.61\\
\\
 6A    & 2007-02-18 & 01:26:25 &  13:59:01 & 12.54\\
 \\
\hline
 \end{tabular}
\end{center}
Note.~---~(1) ATCA configuration;~(2) start date of observation (UT);~(3) and (4) start and end of observations (UT);~(5) time on source.
\end{table*}

\subsection{\hi\ data cubes}
Using the {\sc miriad} software package \citep{MIRIAD} the raw $uv$ data for NGC~1705 were reduced to produce calibrated, deconvolved \hi\ data cubes.  After flagging the first and last five correlator channels the data were split into primary calibrator, secondary calibrator and source subsets.  Corrections were determined for the antenna gains, delay terms and bandpass shapes.  The secondary calibrator was used to determine time-dependent phase and antenna gain corrections.

The calibrated source data were continuum-subtracted by fitting a first-order polynomial to the line-free channels.  The $uv$ data were transformed to the image domain to create a naturally-weighted image cube.  Deconvolution of the cube was carried out using a Steer {\sc clean} algorithm \citep{steer_clean}.  Each channel of the dirty image was {\sc clean}ed down to 2.5 times the typical r.m.s.~flux of a line-free channel or for 50\,000 iterations, whichever condition was met first in practice.  After deconvolution, each of the {\sc clean} components was convolved with a Gaussian approximation of the synthesised beam.  The full width at half maximum of the beam is $16.7''~\times~14.5''$ (412~pc~$\times$~358~pc for $D=5.1$~Mpc).  The spectral resolution of the cube is $dV~=~3.48$~\kms.  No Hanning smoothing was applied.  The noise in a line-free channel is Gaussian distributed with a standard deviation of 0.7~mJy~beam$^{-1}$.  


\subsection{Moment maps}
The Groningen Image Processing SYstem \citep[{\sc gipsy},][]{gipsy}  was used to smooth the \hi\ data cube down to a resolution of $40''~\times~40''$ using a Gaussian convolution function.  A flux cut-off of 2.5 times the r.m.s.~flux of a line-free channel was applied to remove the noise.  This smoothed, flux-cut cube was applied as a mask to the full-resolution cube.  \hi\ moment maps were extracted from the resulting masked cube.  The average intensity of all the pixels in the \hi\ total-intensity map that had a corresponding signal-to-noise ratio in the range $2.75 \leq S/N \leq 3.25$ was used as an intensity cut-off for the map.  The new \hi\ zero-moment map was then used to mask the other moment maps.

\section{\hi\ properties}\label{HI_properties}
\subsection{Channel maps}
Grey-scale representations of the NGC~1705 \hi\ channel maps are presented in Fig.~\ref{high_res_channels}.  The emission exhibits the pattern of a rotating disc, with the southern and the northern parts of the galaxy being blue- and red-shifted, respectively.  The channel maps reveal dense central \hi\ concentrations separated by $\sim 30''$ with more diffuse \hi\ surrounding them.  A drawn out \hi\ feature $\sim 100''$ in angular extent in a north-westerly direction is visible in the channel maps with heliocentric velocity 628.24~\kms~$\lesssim~V_{hel}~\lesssim$~670.09~\kms.  

\begin{figure*}
	\centering
	\includegraphics[width=2\columnwidth, angle=0]{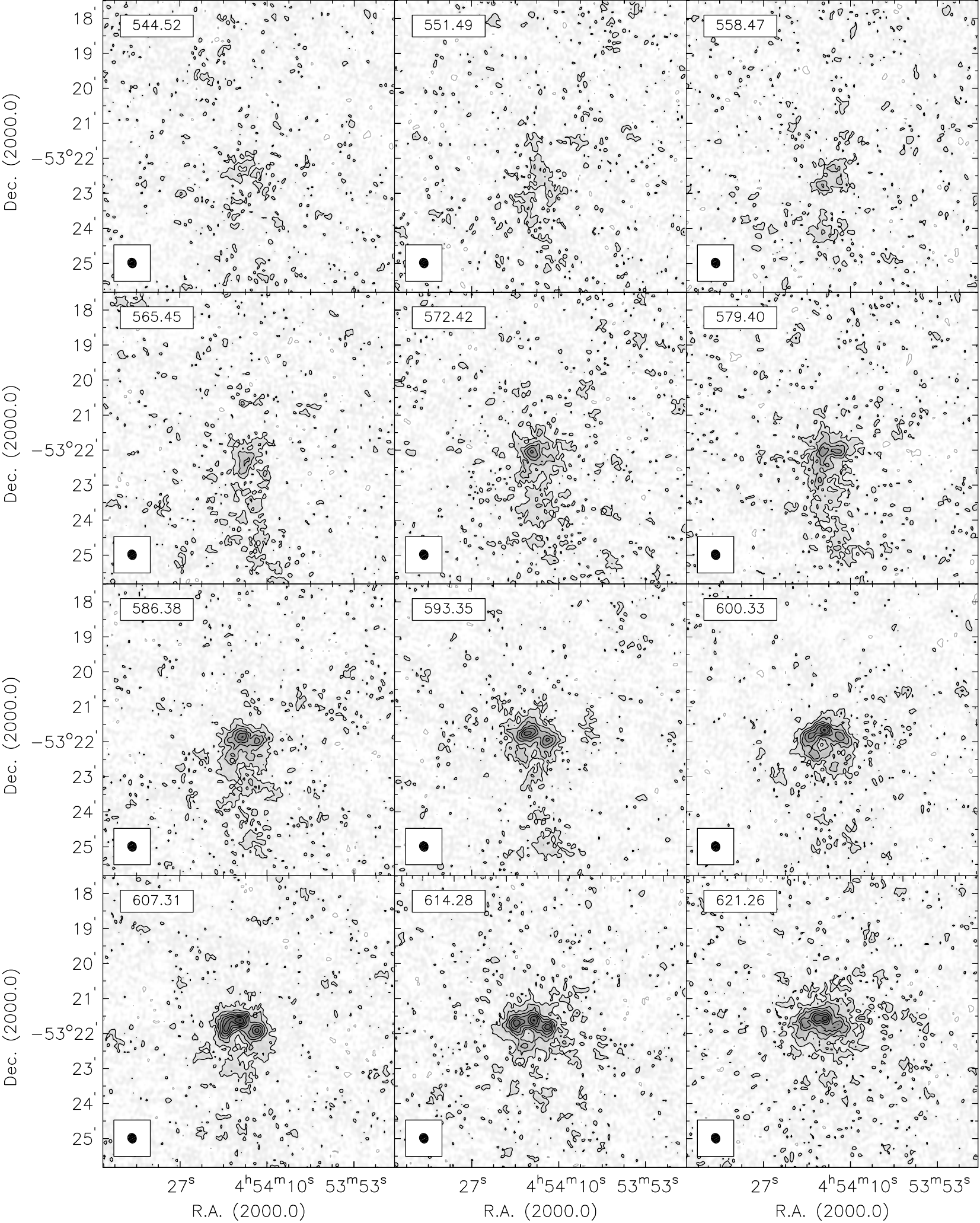}
	\caption{Channel maps of the NGC~1705 \hi\ data cube.  The heliocentric radial velocity (km s$^{-1}$) of each channel is shown in the upper left corner.  The half-power-beam-width ($16.7''~\times~14.5''$) is shown in the bottom left corner.  The greyscale range is from -$\sigma$  to $30\sigma$ where $\sigma=0.7$~mJy~beam$^{-1}$ is the r.m.s. of the noise in a line-free channel.  Contour levels are at -$\sigma$ (grey) and $2-30 \sigma$ in steps of  in steps of 3$\sigma$ (black).}
	\label{high_res_channels}
\end{figure*}
\addtocounter{figure}{-1}

\begin{figure*}
	\centering
 	\includegraphics[width=2\columnwidth, angle=0]{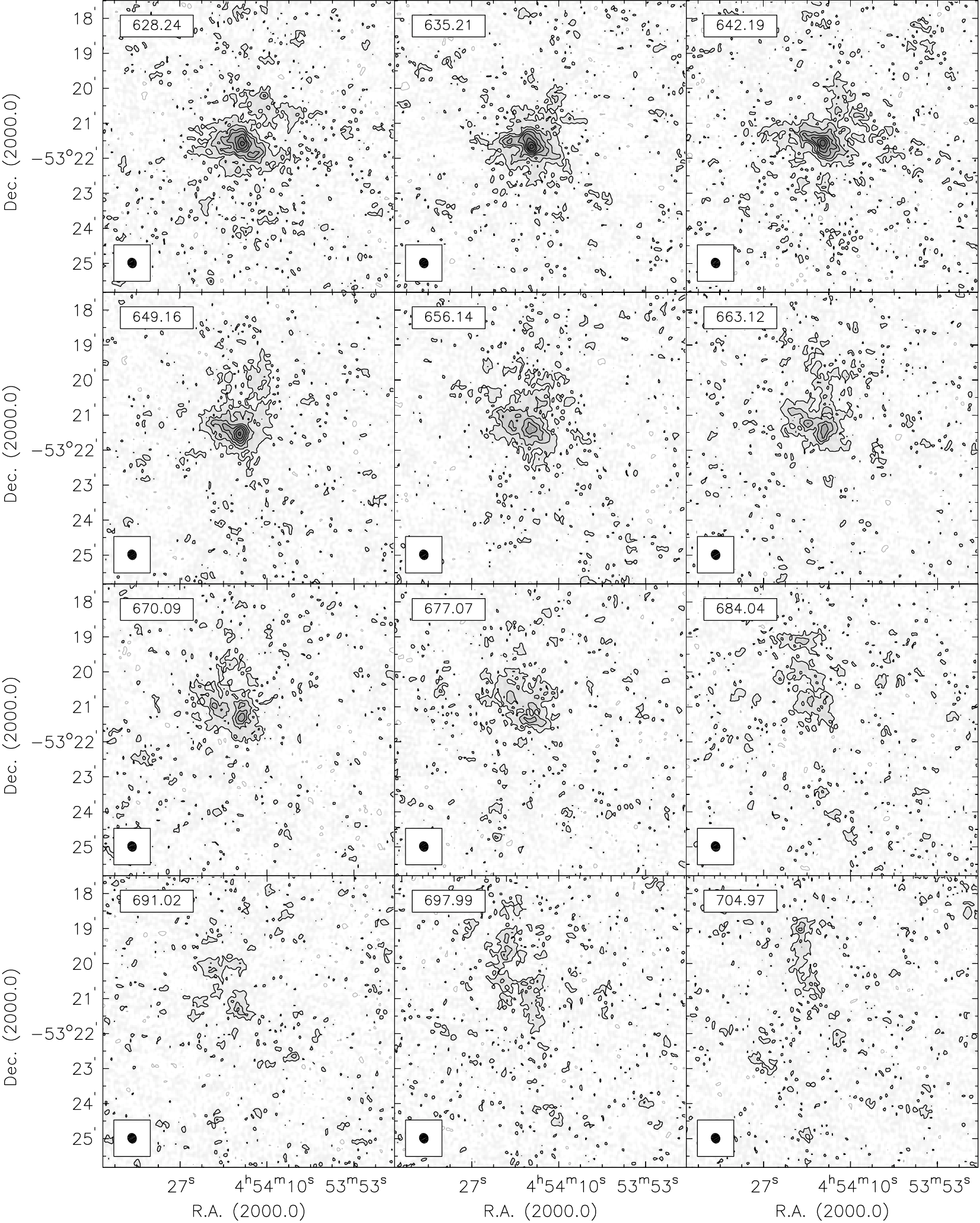}
	\caption[]{Continued.}
\end{figure*}

\subsection{Global profile}\label{1705_global_profile}
\begin{figure}
	\begin{centering}
	\includegraphics[angle=0,width=1.0\columnwidth]{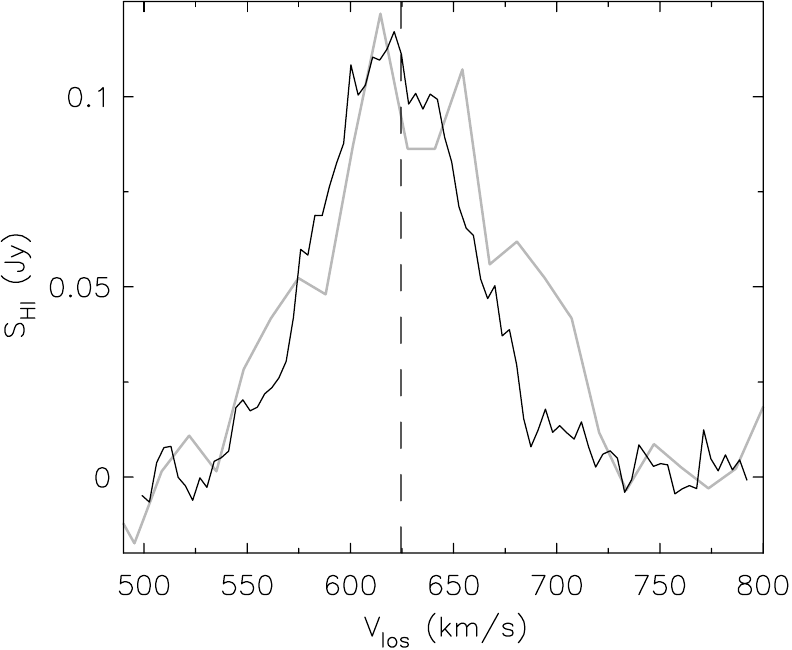}
	\caption[NGC 1705 global \hi\ profiles]{NGC~1705 global H\,{\sc i} profile (black).  The dashed vertical line represents the flux-weighted mean velocity at 624~\kms.  For reference the HIPASS spectrum is shown in grey.}
	\label{1705_global_HI_profiles}
	\end{centering}
\end{figure}

A global \hi\ profile generated by summing the emission in each channel of the \hi\ data cube is presented in Fig.~\ref{1705_global_HI_profiles}.  Given that  the shortest baselines used for the \hi\ observations were of length $\sim~31$~m\symbolfootnote[2]{From the EW352 antenna configuration.} our \hi\ data should be sensitive to \hi\ structures as large as $\sim 23'$.  Since the angular size of NGC~1705 is $\sim 10'$ the amount of \hi\ missed in the global \hi\ profile is expected to be small.  For reference the HIPASS spectrum is also shown in Fig.~\ref{1705_global_HI_profiles}.  It confirms that very little flux is missed by our new observations.

Profile widths  calculated as the differences between the high and low velocities of the galaxy at 20~$\%$ and 50~$\%$ of the peak flux density are W$_{20}=120.2\pm3.5$~\kms\ and W$_{50}=83.5\pm3.8$~\kms, respectively.  The systemic velocity calculated as the flux-weighted mean velocity is $V_{sys}=624.4$~\kms.  The total H\,{\sc i} mass is calculated as 
\begin{equation}
M_{HI}=2.36\times 10^5\times D^2\times\int F dV,
\end{equation}
where $D$ is the distance to the galaxy in units of Mpc and $\int F dV$ is the total H\,{\sc i} line flux in units of Jy~beam$^{-1}$~km~s$^{-1}$.  Our determination of $M_{HI}=2.2\pm0.2\times 10^8$~\msun\ assumes the H\,{\sc i} to be optically thin.  



\subsection{Total intensity map}\label{1705_mom0}
\begin{figure}
	\centering
	\includegraphics[angle=0,width=1.0\columnwidth]{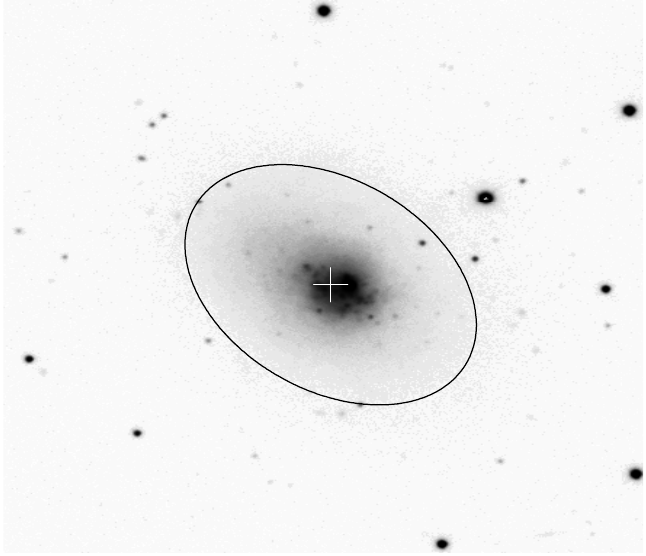}	
	\caption{$R$-band continuum image of NGC~1705.  The black ellipse has a semi-major axis length of 50.1$''$.  This is the isophotal radius within which 90$\%$ of the $R$-band continuum flux is contained.  The white cross marks the position of the photometric centre.}
	\label{SINGG_R_image}
\end{figure}

The \hi\ total intensity (moment zero) map of NGC~1705 extracted from the \hi\ data cube is shown in Fig.~\ref{1705_mom0}.  Also shown is a spatially smoothed version of the  map with resolution 33.2$''~\times~28.6''$.  For comparative purposes an $R$-band continuum image of the stellar disc from the Survey for Ionization in Neutral Gas Galaxies \citep[SINGG,][]{meurer_SINGG_96} is shown in Fig.~\ref{SINGG_R_image}.  To facilitate length-scale comparisons between images an ellipse of semi-major axis length $a=50.1''$ is included in all the maps.  The ellipse represents the radius ($R_{90}$) within which 90$\%$ of the $R$-band flux is contained.


No distinct spiral structure is visible in the \hi\ total intensity maps of NGC~1705.  The high-resolution map shows the \hi\ emission to be concentrated near the optical centre of the galaxy with $\sim 5.1 \times 10^7$~\msun\ of \hi\ (roughly half of the total \hi\ content) residing within $R_{90}$.  Our new \hi\ data resolve this central \hi\ concentration into three over-densities (most clearly seen in the \hi\ channel maps) with a combined mass of $\sim 3\times 10^7$~\msun.  The east- and west-most concentrations have their peaks separated by $\sim35''$ (0.8~kpc), straddling the extremely luminous super star cluster NGC~1705-1 harboured in the galaxy's stellar disc.

\begin{figure*}
	\centering
	\includegraphics[angle=-90,width=2\columnwidth]{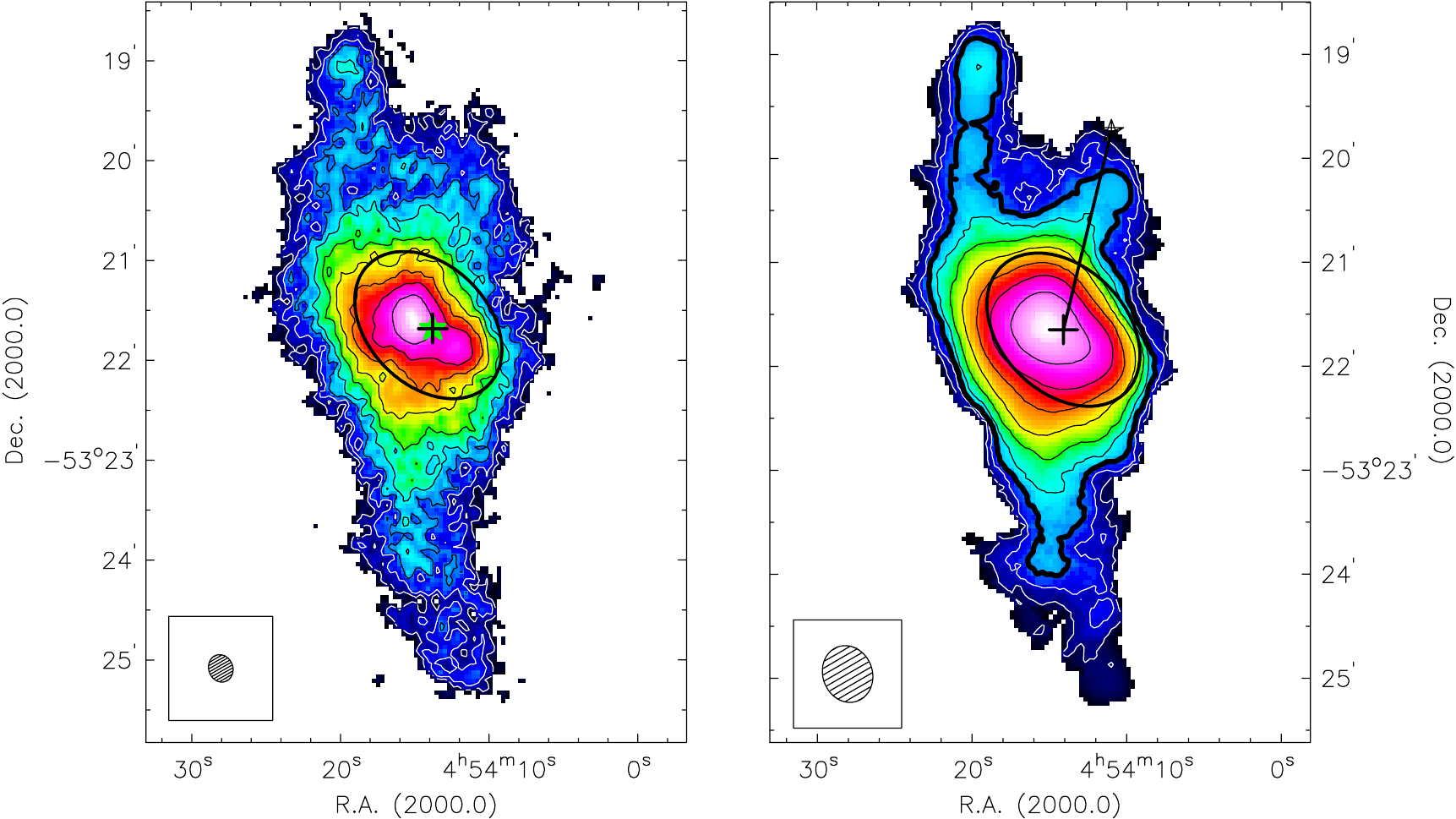}
	\caption{NGC~1705 \hi\ total-intensity maps.  Left panel: high-resolution ($16.7''~\times~14.5''$) map from the \hi\ data cube.  Contour levels are at 7.5, 11.2, 17.0, 25.1, 37.2, 55.0, 83.2, 123.0, 186.2~mJy~beam$^{-1}$~\kms.  The corresponding \hi\ column densities are 1.1, 1.7, 2.1, 4.0, 5.9, 8.7, 13.2, 19.6 and 29.7~$\times 10^{20}$~cm$^{-2}$.  Right panel: low-resolution ($33.2''~\times~28.6''$) map created by convolving the \hi\ data cube with a Gaussian kernel.  Contour levels are at 15.5, 24.0, 36.3, 56.2, 85.1, 131.8, 209.0, 309.0 and 578.6~mJy~beam$^{-1}$~\kms.  The corresponding \hi\ column densities are 0.63, 0.97, 1.4, 2.2, 3.4, 5.3, 8.5, 12.5 and 23.5~$\times 10^{20}$~cm$^{-2}$.  The 24.0~mJy~beam$^{-1}$~\kms\ ($0.97\times 10^{20}$~cm$^{-2}$) contour, marked by the thick black line, is reproduced in Fig.~\ref{SINGG_Ha_image}.  This contour traces the distribution of possibly outflowing \hi.  The black ellipses have a semi-major axis length of 50.1$''$.  This is the isophotal radius within which 90$\%$ of the $R$-band continuum flux is contained.  The black crosses mark the position of the $R$-band photometric centre.  The green-filled star in the left panel marks the central position of the super star cluster NGC~1705-1.  The thick black line in the right-hand panel depicts the approximate length of the extended \hi\ feature discussed in Section~\ref{blow_out}.  The hatched circles in the lower corners of each panel represent the half power beam width of the synthesised beam.}
	\label{1705_mom0}
\end{figure*}

\subsection{Velocity field}\label{1705_mom1}
The intensity-weighted-mean \hi\ velocity field is shown in Fig.~\ref{1705_mom1}.  Small wiggles in the iso-velocity contours are indicative of small-scale streaming motions within the \hi\ disc.  There is no evidence of large-scale perturbations to the velocity field, often associated with the presence of a bar in the mass distribution.  The extended \hi\ feature seen in the low-resolution \hi\ total intensity map (marked by a black line in the right-hand panel of Fig.~\ref{1705_mom0}) has associated line-of-sight velocities consistent with those of the nearby inner portion of the \hi\ disc.  The velocity field as well as the \hi\ channel maps do not suggest the central \hi\ concentrations to be largely separated in velocity from the rest of the disc.  Over all, the system seems to have its kinematics dominated by circular motion.

\begin{figure}
	\centering
	\includegraphics[angle=-90,width=1.0\columnwidth]{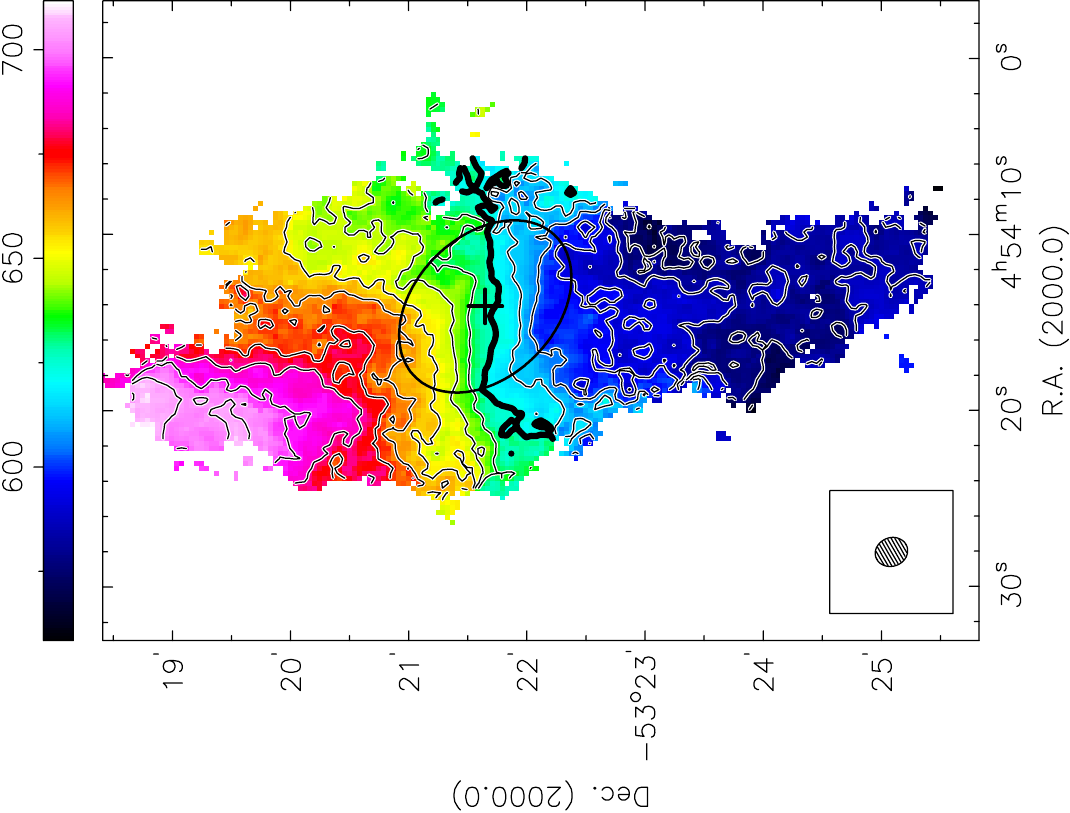}
	\caption{NGC~1705 intensity-weighted-mean \hi\ velocity field.  The southern and northern portions of the galaxy are the approaching and receding sides, respectively.  Contours are separated by 10~km~s$^{-1}$ with the thick contour marking the systemic velocity at 624~km~s$^{-1}$.  The colour scale is specified by the colour bar in units of \kms.  The black ellipse has a semi-major axis length of 50.1$''$.  This is the isophotal radius within which 90$\%$ of the $R$-band continuum flux is contained.  The black cross marks the position of the photometric centre.  The hatched circle in the lower corner represents the half power beam width of the synthesised beam.}
	\label{1705_mom1}
\end{figure}

\subsection{Second-moment map}
\begin{figure}
	\centering
	\includegraphics[angle=-90,width=1.0\columnwidth]{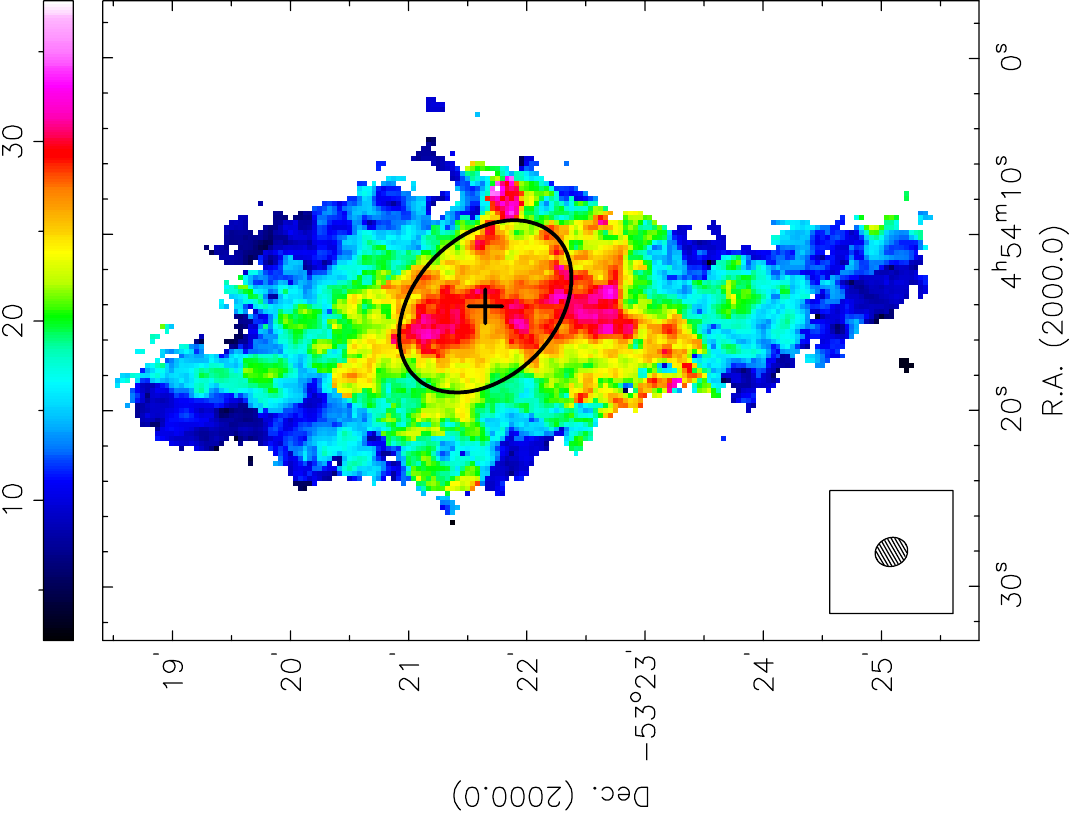}
	\caption{\hi\ second-moment map of NGC~1705.  The colour scale is specified by the colour bar in units of \kms. The black ellipse has a semi-major axis length of 50.1$''$.  This is the isophotal radius within which 90$\%$ of the $R$-band continuum flux is contained.  The black cross marks the position of the photometric centre.  The hatched circle in the lower corner represents the half power beam width of the synthesised beam.}
	\label{1705_mom2}
\end{figure}

\begin{figure}
	\centering
	\includegraphics[angle=0,width=\columnwidth]{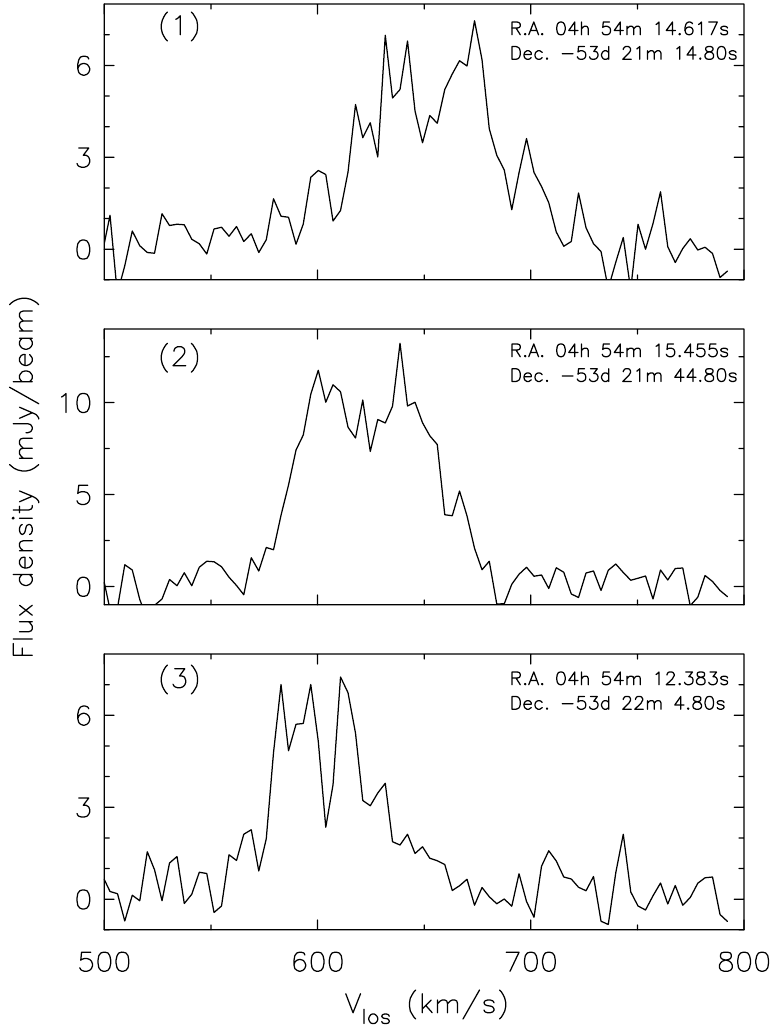}	
	\caption{Line profiles extracted from the \hi\ data cube of NGC 1705 at the positions marked by the small white crosses in Fig.~\ref{SINGG_Ha_image}.  Profiles 1, 2 and 3 correspond to the top, middle and bottom crosses, respectively.  }
	\label{hilineprofiles}
\end{figure}

The second-moment map of NGC~1705 is shown in Fig.~\ref{1705_mom2}.   A very sharp rise in second-moments is observed from the outer to the inner \hi\ disc, with typical central second-moments of the order of $\sigma_{HI}\sim 30$~\kms.  These broad line profiles are not expected to be representative of the thermal velocity dispersion of the \hi.  \citet{young_et_al_2003} analysed the shapes and widths of \hi\ line profiles of three dwarf galaxies observed with the Very Large Array, finding higher star formation rates to be correlated with higher fractions of asymmetric and double-peaked \hi\ line profiles.  They interpret the result as evidence of the \hi\ being stirred up by the mechanical energy being deposited into the ISM by high-mass star formation.  Figure~\ref{hilineprofiles} shows \hi\ line profiles extracted from the \hi\ data cube of NGC~1705 at the positions marked by small crosses in Fig.~\ref{SINGG_Ha_image}.  These line profiles clearly contain more than a single  kinematic component, indicative of a highly dispersed, perhaps expanding, \hi\ component that is being driven by stellar feedback processes. 


\section{A galactic wind blow-out?}\label{blow_out}
The low-resolution \hi\ total intensity map of NGC~1705 reveals a drawn-out \hi\ feature emanating from the centre of the galaxy in a north-westerly direction.  If measured from the photometric centre of the galaxy this feature has an angular extent of $\sim 130''$ (3.3~kpc, black line in the right-hand panel of Fig.~\ref{1705_mom0}).  \citet{meurer_1705_2} treat it as evidence for a galactic wind blow-out.  They propose the \hi\ spur to be a result of the ambient inter-stellar medium (ISM) that has been swept up by an expanding, over-pressurised bubble of thermalised supernovae remnants associated with the starburst.  

\begin{figure}
	\centering
	\includegraphics[angle=0,width=\columnwidth]{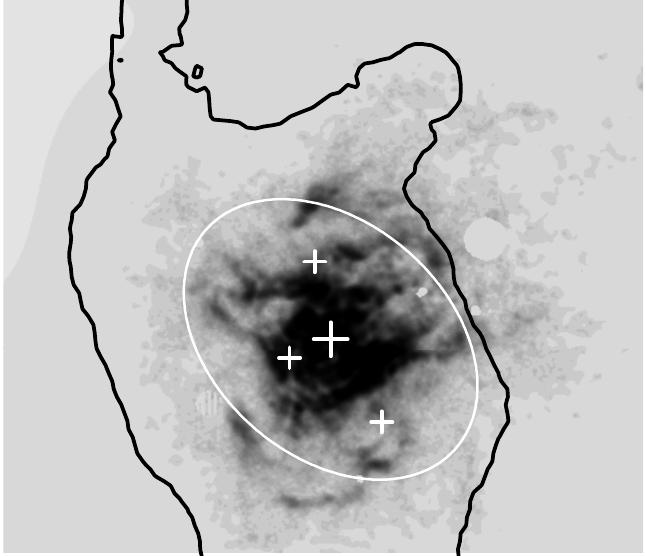}	
	\caption{Narrow-band \Ha\ image of NGC~1705.  The black line marks the $1.0\times 10^{20}$~cm$^{-2}$ \hi\ column density contour.  This contour, also shown in the right-hand panel of Fig.~\ref{1705_mom0}, traces the distribution of the \hi\ spur.  Note how some of the \Ha\ arcs are coincident with the \hi\ spur.  The white ellipse has a semi-major axis length of 50.1$''$.  This is the isophotal radius within which 90$\%$ of the $R$-band continuum flux is contained.  The large white cross marks the position of the photometric centre.  The smaller white crosses, from top to bottom, mark the respective positions at which line profiles 1, 2, and 3 shown in Fig.~\ref{hilineprofiles} were extracted.}
	\label{SINGG_Ha_image}
\end{figure}

An \Ha\ narrow-band image of NGC~1705 produced using SINGG data (Fig.~\ref{SINGG_Ha_image}) reveals multiple arcs and loops in the \Ha\ distribution that extend well beyond the $R$-band $R_{90\%}$ radius.  Clearly evident is a component of the \Ha\ emission coincident with the \hi\ spur.  The one-sided nature of the outflows is predicted by models of galactic winds \citep[e.g.][]{maclow_1999} in which the energy source is displaced from the disc plane.  It is conceivable that  portions of both the \hi\ and \Ha\ mass components in NGC~1705 are entrained in the same galactic-scale blow-out that is removing material from the galaxy's ISM into its halo.  

\section{Baryonic mass distributions and rotation curve}
In this section we measure the radial distributions of \hi\ and stellar mass in NGC~1705.  In combination with the rotation curve these profiles allow us to study the radial distribution of dark matter within the galaxy.

\subsection{Radial distribution of \hi.}
\begin{figure}
	\begin{centering}
 	\includegraphics[width=1.0\columnwidth]{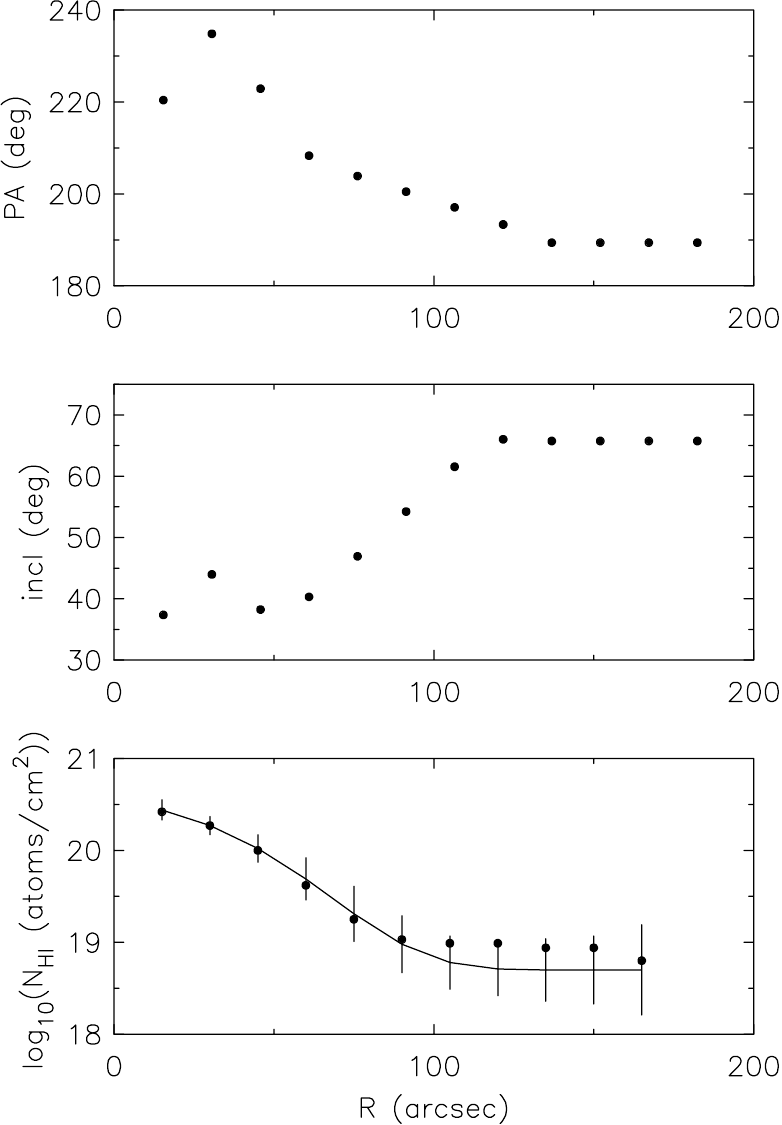}
	\caption{Radial profiles of the position and inclination angles (top and bottom panels, respectively) of thin rings extracted from the \hi\ column density map of NGC~1705.  The bottom panel shows the azimuthally-averaged \hi\ column density within each ring (black-filled circles).  Error bars represent the r.m.s. spreads of the \hi\ column densities in the rings.  The black curve represents a Gaussian fit to the data.}
	\label{PA_INCL_NHI_radial_profiles}
	\end{centering}
\end{figure}

A map of the \hi\ column densities ($N_{HI}$) in NGC~1705 was generated from the \hi\ total intensity map shown in Fig.~\ref{1705_mom0}.  The $N_{HI}$ map was divided into rings.  Ellipses were fit to the rings in order to determine the radial variations of their inclinations and position angles.  The resulting inclination and position angle radial profiles are shown in the top two panels of Fig.~\ref{PA_INCL_NHI_radial_profiles}.  The \hi\ column densities within each ring were azimuthally averaged to produce an $N_{HI}$ radial profile, shown in the bottom panel of Fig.~\ref{PA_INCL_NHI_radial_profiles}.

\subsection{Radial distribution of stellar mass}\label{stellar_ML}
NGC~1705 has long been known to harbour an extended distribution of old stars and a more compact distribution of young stars \citep{meurer_1705_1992}.  By comparing synthetic colour-magnitude diagrams to Hubble Space Telescope optical and near-infrared photometry, \citet{Annibali_2003} estimate the total mass contained in stars younger than 1~Gyr to be approximately $6\times 10^7$~\msun, while $2.2\times 10^8$~\msun\ is contained within the older stellar population.  We use these mass estimates together with radial flux profiles extracted from \emph{Spitzer} 3.6~\micron\ and GALEX far-ultraviolet imaging to derive the radial distribution of mass surface density for each stellar population.

\begin{figure}
	\begin{centering}
	\includegraphics[width=1\columnwidth]{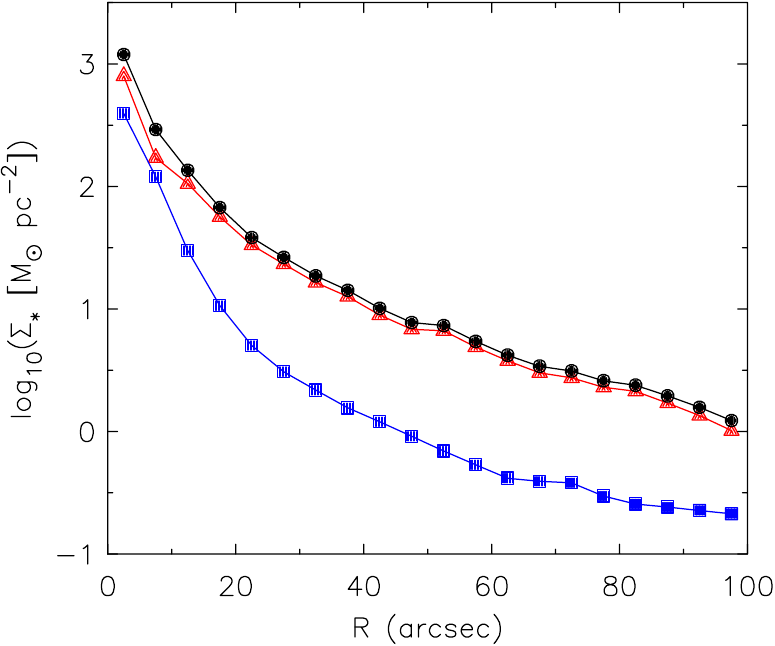}
	\caption{Radial distributions of stellar mass surface density for 1) the old ($>$ 1 Gyr) stellar population (red-filled triangels), 2) the young stellar ($<$ 1 Gyr) population (blue-filled squares), and 3) the total stellar population (black-filled circles).}
	\label{FUV_IR_profiles}
	\end{centering}
\end{figure}

The far-ultraviolet and 3.6~\micron\ images were divided into sets of concentric rings of width 5~arcsec and with centre position $\alpha_{2000}$~=~04$^{\mathrm{h}}$~54$^{\mathrm{m}}$~13.$^{\mathrm{s}}$532, $\delta_{2000}$~=~$-53^{\circ}$~21$'$~39.26$''$.  Assuming a distance $D=5.1$~Mpc for NGC~1705, the azimuthally-averaged flux density of each ring was converted to a mass surface density in units of \msun~pc$^{-2}$.  Figure~\ref{FUV_IR_profiles} shows these profiles for the old (3.6~\micron\ imaging) and young (far-ultraviolet imaging) stellar populations as red-filled triangles and blue-filled squares, respectively.  The radial distribution of mass surface density for the total stellar population is represented in Fig.~\ref{FUV_IR_profiles} by the black-filled circles.  Although the 3.6~\micron\ imaging of NGC~1705 is dominated by emission from old stars, some emission will be contributed by young stars.  Similarly, old stars will produce some far-ultraviolet emission.  It is for this reason that the total-stellar-mass-surface-density curve shown in Fig.~\ref{FUV_IR_profiles} (black-filled circles) serves as an upper limit for the total amount of mass contained within stars in NGC~1705.

\subsection{Rotation curve}\label{rotation_curve_sec}
Despite the possibility of a large-scale mass outflow in NGC~1705, the  overall \hi\  kinematics of the system are still dominated by  rotation.  The velocity field therefore serves as a useful tracer of the kinematics that can be used to extract a rotation curve.  A standard method for deriving a rotation curve of a disc galaxy of intermediate inclination is to fit a tilted ring model \citep{Rogstad1974} to its velocity field.  The method involves modeling the disc as a set of concentric rings within which the gas moves along circular orbits about the dynamical centre of the galaxy.  Each ring is defined by a set of parameters: central coordinates $(x_c, y_c)$, inclination ($i$), position angle ($PA$, in the plane of the sky), systemic velocity of the galaxy ($V_{sys}$), and rotation velocity ($V_{rot}$).  In the case that only circular velocities are considered the line-of-sight velocity ($V_{los}$) at a position ($x, y$) on the sky is given by
\begin{equation}
V_{los}(x,y) = V_{sys}+V_{rot}\sin i\cos\theta,
\end{equation}
where $\theta$ specifies the position angle in the galaxy plane.  $\theta$ and $PA$ are related as follows:

\begin{eqnarray}
\cos\theta&=& {-(x-x_c)\sin PA + (y-y_c)\cos PA\over R},\\
\sin\theta&=& {-(x-x_c)\cos PA + (y-y_c)\sin PA\over R\cos i}.
\end{eqnarray}

\begin{figure}
	\begin{centering}
 	\includegraphics[width=1.0\columnwidth]{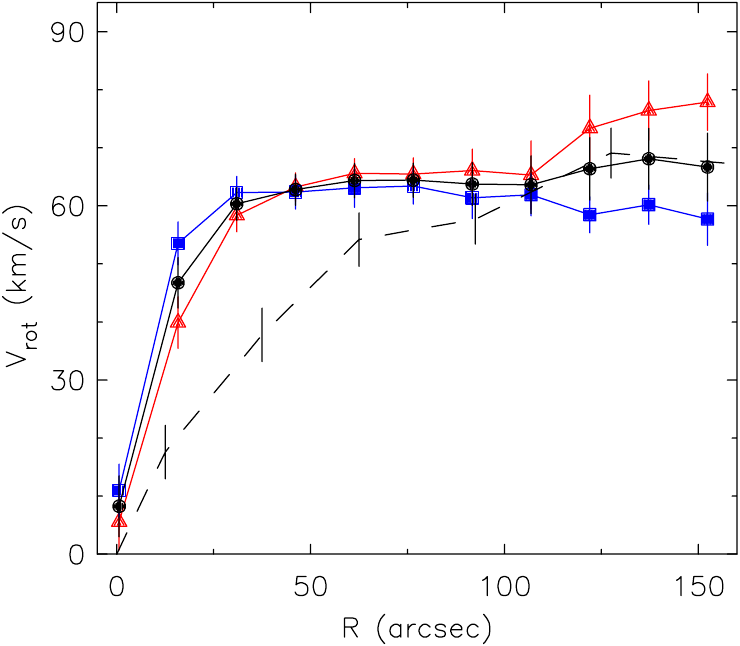}
	\caption{Rotation curves derived for the entire \hi\ disc of NGC~1705 (black-filled circles) as well as the approaching (blue-filled squares) and receding (red-filled triangles) halves of the galaxy by fitting a tilted ring model to the \hi\ velocity field.  Error bars represent the r.m.s. spread of velocities in a given ring.  For comparison the rotation curve derived by \citet{meurer_1705_2} is shown as a dashed curve.}
	\label{vrot_compare}
	\end{centering}
\end{figure}

\begin{figure}
	\begin{centering}
 	\includegraphics[width=1.0\columnwidth]{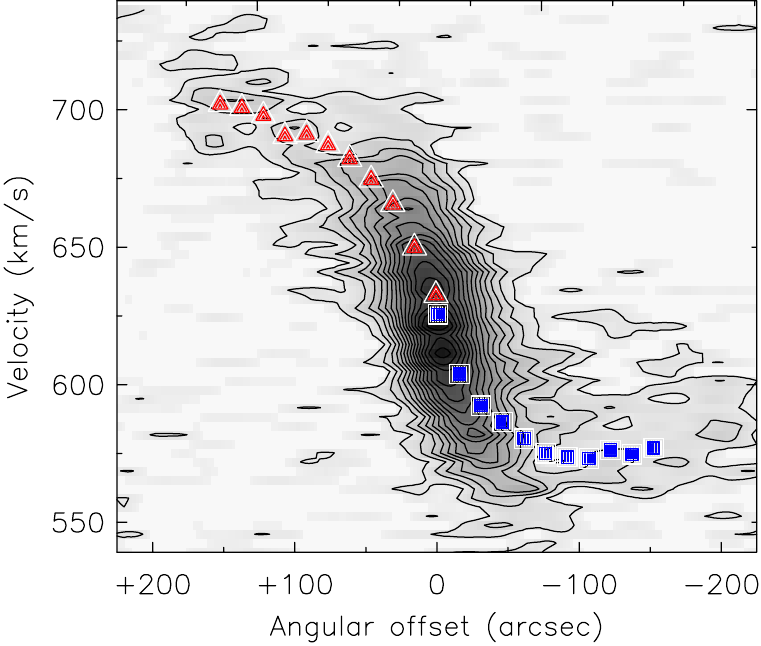}
	\caption{Rotation curves derived for the approaching (blue-filled squares) and receding (red-filled triangles) halves of NGC~1705 overlaid on a position-velocity slice extracted from the \hi\ data cube along the \hi\ kinematic major axis.}
	\label{vrots_pv_overlay}
	\end{centering}
\end{figure}

The $PA$ and $i$ profiles extracted from the $N_{HI}$ map were used to specify the orientations of the rings in our tilted ring model.  The average position of the centres of the ellipses fit to the $N_{HI}$ map was used as the centre of all rings.  Although $V_{sys}=624$~\kms\ is the systemic velocity for NGC~1705 derived from its global \hi\ profile, our best-fitting tilted ring models were generated using $V_{sys}=630$~\kms.  Rotation curves were derived for the approaching and receding halves as well as the entire galaxy (Fig.~\ref{vrot_compare}).   Also shown in Fig.~\ref{vrot_compare} is the rotation curve derived by \citet{meurer_1705_2}.  Our rotation curves clearly rise much more steeply at inner radii than their curve.  The difference can be attributed to our new \hi\ data which are of higher spatial resolution.  In Fig.~\ref{vrots_pv_overlay} we overlay the receding- and approaching-side rotation curves on a position velocity slice extracted from the \hi\ data cube along the kinematic major axis of the \hi\ disc.  Our rotation curves serve as good representations of the rotational dynamics of NGC~1705.

\section{Mass modeling}\label{mm}

The rotation curves of the individual mass components of a galaxy, when summed in quadrature, yield the square of the total rotation velocity, $V_{tot}$.  For stellar, gaseous and dark matter components
\begin{equation}
V_{tot}^2=V_{gas}^2+ V_*^2+V_{DM}^2,
\label{vtot} 
\end{equation}
where $V_{gas}$, $V_*$ and $V_{DM}$ are the dynamical contributions to the total rotation curve of the gas, the stars and the dark matter, respectively.  To account for the presence of Helium and other metals in our mass models we boosted the contribution of $V_{gas}$ by a factor of 1.37. 

While $V_{gas}$ and $V_*$ are based on observations, an analytic form for the dark matter distribution is needed in order to yield a $V_{DM}$ parameterisation that can be used for the mass modeling.  Numerical simulations of the hierarchical growth of Cold Dark Matter (CDM) suggest the existence of a universal form of the equilibrium density profiles of dark matter haloes \citep{NFW_1997}.  What is now commonly referred to as the NFW profile is parameterised as
\begin{equation}
{\rho(r)_{NFW}\over \rho_{crit}}= {\delta_c \over {(r/r_s)(1+r/r_s)^2}}
\end{equation}
where $\delta_c$ is a measure of the density of the Universe at the time of collapse of the dark matter halo, $r_s$ is the characteristic scale radius and $\rho_{crit}=3H^2/8\pi G$ is the critical density of the Universe required for closure.  At small radii this profile scales as $r^{-1}$, thereby predicting extremely steep inner density profiles of dark matter halos known as cusps.  This mass distribution results in a dark matter halo rotation curve of the form
\begin{equation}
V_{NFW}(r) = V_{200}\left[{\ln(1+cx)-cx/(1+cx)\over x[\ln(1+c)-c/(1+c)]}\right]^{0.5}. 
\end{equation}
Given $r_{200}$ as the radius at which the density contrast relative to the critical density of the Universe equals 200, $V_{200}$ is the characteristic velocity at $r_{200}$ and $x=r/r_{200}$.  The concentration parameter $c=r_{200}/r_s$.

Observational determinations of the dark matter density profiles of nearby galaxies often show them to be too shallow to be fit by the theoretical NFW profile \citep{moore_1994,flores_primack,deblok_mcgaugh_1996,marchesini_2002, THINGS_deblok}.  Rather than being cuspy, the dark matter density is found to remain approximately constant at inner radii.  A common empirically-motivated parameterisation of the dark matter density profile is that of a pseudo iso-thermal sphere:
\begin{equation}
\rho(r)_{ISO}=\rho_0\left(1+\left({r\over r_c}\right)^2\right)^{-1},
\end{equation}
where $\rho_0$ is the central dark matter density and $r_c$ is the core radius within which the density remains constant.  The rotation curve corresponding to this mass distribution is
\begin{equation}
V_{ISO}(r)= \left[  4\pi G\rho_0 r_c^2\left(1-{r_c\over r}\arctan{r\over r_c}\right)\right]^{1/2},
\end{equation}
where $G$ is the gravitational constant.

Mass models for a pseudo-isothermal sphere and an NFW halo were fit to the rotation curves of both the approaching and receding halves of the galaxy.  Our parameterisations of the \hi\ distribution (Fig.~\ref{PA_INCL_NHI_radial_profiles}, bottom pannel) and the stellar mass distribution (Fig.~\ref{FUV_IR_profiles}, black-filled circles) were converted to rotation curves using the {\sc gipsy} task {\sc rotmod} \citep{rotmod}.  The {\sc rotmas} task was used to generate mass models by subtracting the \hi\ and stellar rotation curves from the observed rotation curve and then fitting a dark matter rotation curve to the residual.  


\subsection{Results and discussion}
\begin{figure*}
	\begin{centering}
	\includegraphics[angle=0,width=2.1\columnwidth]{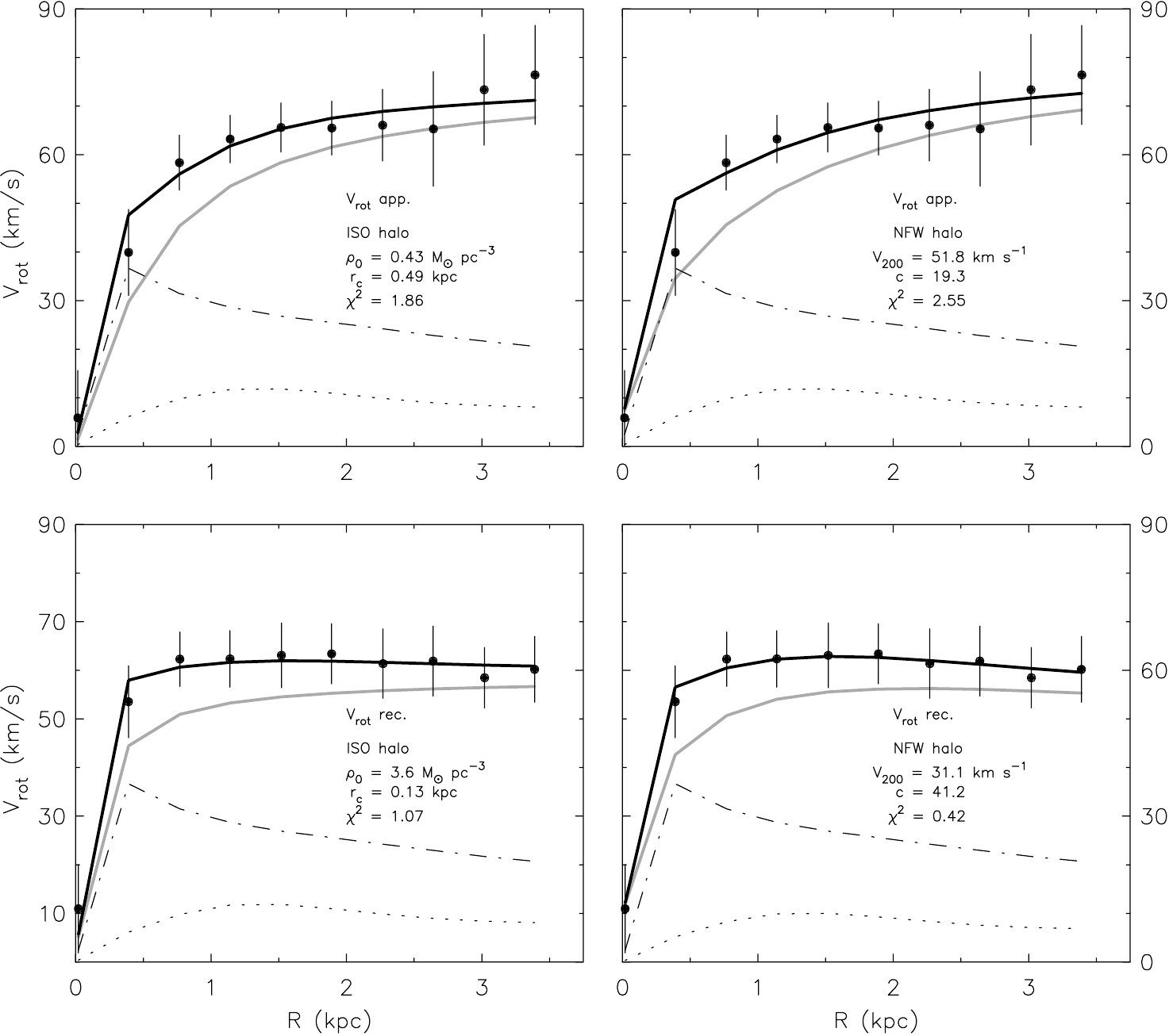}
	\caption[NGC~1705 mass models]{NGC~1705 mass models for pseudo-isothermal sphere (left-hand panels) and NFW halo (right-hand panels) parameterisations of the dark matter halo.  Separate mass models were fit to the rotation curves of the approaching (top panels) and receding (bottom panels) halves of the galaxy.  In all panels the black-filled circles represent the observed rotation curve while error bars represent the uncertainty.  The black dot-dash and dotted curves show the dynamical contributions of the stars and gas to the total rotation curve, respectively.  The solid grey curve shows the dark matter contribution.  The solid black curve represents the total rotation curve.  In each panel the $\chi^2$ goodness-of-fit statistic is presented together with the best-fitting halo parameters.}
	\label{1705_mm_results}
	\end{centering}
\end{figure*}

\begin{figure}
	\begin{centering}
	\includegraphics[angle=0,width=\columnwidth]{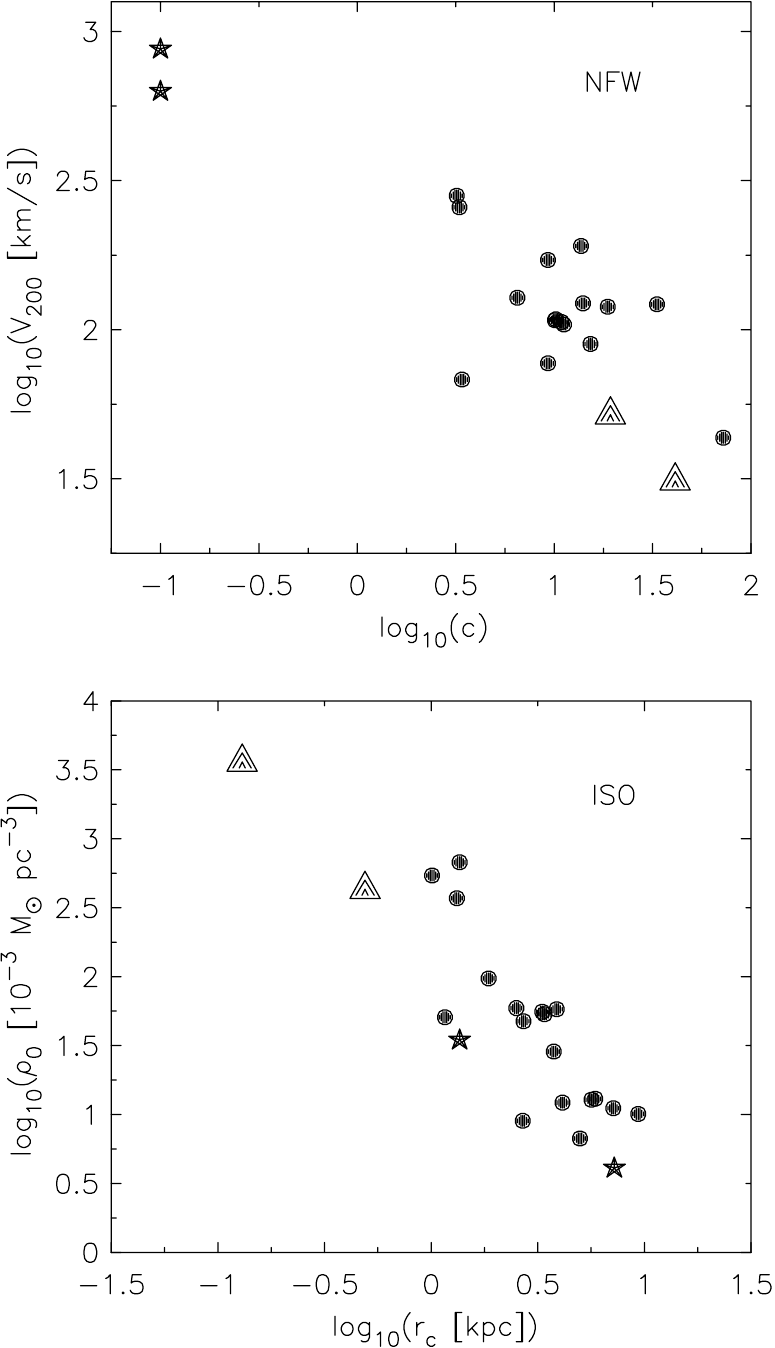}
	\caption{NGC~1705 mass modeling results (triangles) compared to other late-type galaxies from the THINGS survey (circles) as well as NGC~2366 and IC~2574 from \citet{THINGS_oh} (stars).  The upper and lower panels correspond to the NFW and pseudo-isothermal sphere dark matter halo parameterisations, respectively.}
	\label{1705_THINGS_halos}
	\end{centering}
\end{figure}

The mass modelling results are presented in Table~\ref{1705_mass_modeling_results}.  Figure~\ref{1705_mm_results} shows the dynamical contributions of the gas, stars and dark matter to the total rotation curve.  Figure~\ref{1705_THINGS_halos} shows the dark matter halo parameters for each of our four derived mass models compared to the halo parameters of the THINGS (The \hi\ Nearby Galaxy Survey)galaxies.  

\begin{table*}
\begin{center}
\caption{Mass modeling results for NGC~1705.}
\label{1705_mass_modeling_results}
\begin{tabular}{cccccc}
\hline
\hline
\\
\\
		&   										& 	$V_{rot}^{app}$ 	&  	$V_{rot}^{rec}$	 	& 	$V_{rot}^{app}$		&	$V_{rot}^{rec}$		\\
		&										&	ISO				&	ISO					&	NFW				&	NFW				\\
		&									       	&	1 				& 	2 					&	3				&	4				\\
\\
\hline
\\
1		&	$\chi^2	$							&	1.86				&	1.07					&	2.55				&	0.42				\\

2		&	r.m.s. [km s$^{-1}$] 						& 	3.79				&	2.48					&	4.34				&	1.3				\\


3		& $\rho_0$ [M$_{\odot}$ pc$^{-3}$] 				& 	$0.43\pm0.1$		&	$3.6\pm 1.8$			&	...				&	...				\\

4		& $r_c$ [kpc] 								& 	$0.49\pm0.1$		&	$0.13\pm 0.03$		&	...				&	...				\\

5		& $V_{200}$ [\kms] 							& 	...				&	...					&$51.8\pm8.6$			&	$31.1\pm0.8$		\\

6		& $c$ 		 							& 	...				&	...					& $19.3\pm4.2$		&	$41.2\pm2.9$		\\

\\
		&At last measured point on rotation curve:\\
\\
7		& M$_{tot}$ [10$^{9}$ \msun] 					&	4.0				&	2.9					&	4.1				&	2.8				\\

8		& M$_{HI}$/M$_{tot}$ [$10^{2}$] 				& 	1.3				&	1.7					&	1.2				&	1.3				\\

9		& M$_{DM}$/M$_{tot}$ [$10^{2}$]				& 	90.3				&	86.6					&	90.7				&	86.0				\\

10		& M$_{DM}$/M$_{bary}$ 						&	9.3				&	6.5					&	9.7				&	6.4				\\

11		& M$_{tot}$/L$_B$ [M$_{\odot}$/L$_{\odot}$] 		& 	14.8				&	10.8					&	15.4				&	10.4				\\
\\
\hline
 \end{tabular}
\end{center}
\begin{flushleft}
Notes~---~(1)~$\chi^2$ goodness-of-fit statistic;~(2) r.m.s. difference between observed and total rotation curves;~(3)~pseudo-isothermal sphere central density;~(4)~pseudo-isothermal sphere core radius;~(5)~circular rotation speed at virial radius of NFW halo;~(6)~concentration parameter for NFW halo;~(7)~dynamical mass;~(8)~\hi-to-total-mass ratio;~(9)~dark-to-total-mass ratio;~(10)~dark-to-baryonic-mass ratio;~(11)~total-mass-to-$B$-band-light ratio.  
\end{flushleft}
\end{table*}

Our best-fitting mass models are for the receding half of the galaxy.  Both the pseudo-isothermal sphere and NFW dark matter halo parameterisations show the galaxy to be dark-matter dominated at all radii.  This is true even for the innermost portion of the galaxy where the dense concentration of stellar mass largely contributes to the rotational dynamics.  From these models the total (dynamical) mass of the galaxy is measured to be $\sim 2.8\times 10^9$\msun.  Roughly $86\%$ of the total mass is constituted by dark matter, while the average ratio of dark-to-baryonic mass is $\sim 6.4$.  Our best-fitting pseudo-isothermal sphere parameterisation for the dark matter halo suggests the dark matter to be concentrated as densely as $\rho_0=3.6\pm 1.8$~\msunppc\ within a core of radius $r_c=0.13\pm 0.03$~kpc.  This estimate for the central dark matter density if much higher than the estimate made by  \citet{meurer_1705_2} who also parameterise the dark matter halo as a pseudo-isothermal sphere, finding $\rho_0\approx 0.1$~\msunppc.  The large difference can be mainly attributed to our derived rotation curve which rises steeply at inner radii.  The \citet{meurer_1705_2} rotation curve, based on lower resolution \hi\ data, rises much more slowly with radius.  Our models require much more dark matter near the centre of the galaxy to produce the observed rotation speeds.  

Our NFW halo parameterisation for the receding half of the galaxy yields the best-fitting mass model with $\chi^2=0.42$.  The model produces a total rotation curve that fits both the sharply rising inner portion and the gradually declining outer portion of our derived rotation curve.  The estimated concentration parameter $c=41.2\pm 2.9$ is higher than most of the late-type THINGS galaxies mass modelled by \citet{THINGS_deblok}.  In contrast, NGC~1705 has a $V_{200}$ parameter much lower than all of the THINGS galaxies.  

Our mass models for the approaching half of NGC~1705 are not as well fitted as those for the receding half.  For both the pseudo-isothermal sphere and NFW halo parameterisations the inner portion of the rotation curve is over-estimated.  Despite implying a total mass of $\sim 4\times 10^9$~\msun, the models show the dark matter not to dominate the dynamics at all radii.  At radii $R\lesssim 0.5$~kpc it is the stellar disc that almost completely dominates the gravitational potential.  Regarding the radial distribution of dark matter, our pseudo-isothermal sphere model suggests there to be a dark matter core of radius $r_c=0.49\pm 0.1$~kpc and  density $\rho_0=0.43\pm 0.1$~\msunppc.  These parameters are in better agreement with those of \citet{meurer_1705_2} yet still make the dark matter core of the galaxy significantly more dense and compact than other late-type systems (Fig.~\ref{1705_THINGS_halos}).  The best-fitting NFW halo for the approaching half of the galaxy is similar to that of the receding half - again suggesting a high concentration parameter together with a low $V_{200}$.  All of the mass modeling results taken together imply NGC~1705 to contain a lot of dark matter with a dense, compact core.  The extreme density and compactness of the inner halo of NGC~1705 make it comparable to early-type dwarf spheroidals which have typical dark matter core densities of $\rho_0\sim 0.5$~\msun~pc$^{-3}$ \citep{gilmore_2007}.

\section{Summary and Conclusions}\label{1705_HI_summary}
We have used \hi\ synthesis data for NGC~1705 to study the properties of its extended gaseous disc.  We have produced high-quality \hi\ channel and moment maps.  A total \hi\ mass of $2.2\pm 0.2\times 10^8$~\msun\ is estimated from the global \hi\ profile.  A rotation curve for the galaxy is derived by fitting a tilted ring model to the \hi\ velocity field.  The rotation curve rises steeply near the centre of the galaxy and then levels out to a roughly constant velocity in the outer disc.  The galaxy is kinematically lopsided with different asymptotic velocities derived for each side of the \hi\ disc.  The \hi\ and stellar contributions to the rotation curve are estimated using our new \hi\ data together with \emph{Spitzer} 3.6~\micron\  and GALEX far-ultraviolet imaging, respectively. 

The high spatial resolution of our \hi\ data allows the inner \hi\ distribution to be resolved into at least two over-densities which straddle the system's extremely luminous central super star cluster, NGC~1705-1.  This star cluster is the main energy source powering a galactic outflow which is revealed in our channel and moment maps as a drawn-out \hi\ feature extending from the centre of the galaxy in the same direction as an extended \Ha\ component.  The structure of the intensity-weighted-mean \hi\ velocity field is that of a rotating disc.

The dark matter halo of the galaxy is parameterised as a pseudo-isothermal sphere and an NFW halo.  All mass models suggest the the mass budget of the galaxy to be dominated by dark matter.  Models for the receding half of the galaxy are consistent with an extremely dense and compact dark matter core.  In these models it is the dark matter that dominates the gravitational potential at all radii.  The mass models for the approaching half of the galaxy are less well-fitted, yet are still consistent with the observations.  These models show the stellar mass to dominate the gravitational potential at inner radii ($\lesssim 0.5$~kpc), yet still measure the core to be compact ($rc=0.49\pm 0.1$~kpc) and dense ($\rho_0=0.43\pm 0.1$~\msunppc).  The best-fitting parameters for all of our pseudo-isothermal sphere and the NFW halos are more extreme than those of almost all the THINGS galaxies.

\section{Acknowledgements}\label{acknowledgements}
The work of E.C.E. is based upon research generously supported by the South African SKA project.  All authors acknowledge funding received from the South African National Research Foundation.  The work of W.J.G.deB. is based upon research supported by the South African Research Chairs Initiative of the Department of Science and Technology and the National Research Foundation.  The Australia Telescope Compact Array is part of the Australia Telescope which is funded by the Commonwealth of Australia for operation as a National Facility managed by CSIRO.  This work is based [in part] on observations made with the \emph{Spitzer} Space Telescope, which is operated by the Jet Propulsion Laboratory, California Institute of Technology under a contract with NASA.


\end{document}